\documentclass[twocolumn,twoside,slac_two]{revtex4}
\usepackage{graphicx}
\usepackage{fancyhdr}
\usepackage{graphics}
\usepackage{epstopdf}
\usepackage{textpos}
\newcommand{\met}       {\mbox{$\not\!\!E_T$}}
\newcommand{\mcatnlo}    {\mbox{\textsc{mc@nlo}}}
\begin{document}

\title{\centering Top Quark Properties}
\author{
\centering
\begin{center}
Yvonne Peters \\ (On Behalf of the ATLAS, CDF, CMS and D0 Collaborations)
\end{center}}
\affiliation{\centering University of Manchester, United Kingdom}
\begin{abstract}
Since its discovery in 1995 by the CDF and D0 collaborations at the
Fermilab Tevatron collider, the top quark has undergone intensive
studies. Besides the Tevatron experiments, with the start of the LHC in
2010 a top quark factory started its operation. It is now possible  to
measure top quark properties simultaneously at four different experiments,
namely ATLAS and CMS at LHC and CDF and D0 at Tevatron. 
Having collected thousands of top quarks each, several top quark
properties have been measured precisely, while others are being measured for
the first time. In this article, recent measurements of top quark
properties from ATLAS, CDF, CMS and D0 are presented, using up to 5.4
fb$^{-1}$ of integrated luminosity  at the Tevatron and 1.1 fb$^{-1}$ at the LHC. In particular,
measurements of the top quark mass, mass difference, foward backward
charge asymmetry, $t\bar{t}$ spin correlations, the ratio of branching
fractions, $W$ helicity, anomalous couplings, color flow and the
search for flavor changing neutral currents are discussed. 
\end{abstract}

\maketitle
\thispagestyle{fancy}

\section{Introduction}
The heaviest elementary particle known today is the top quark, with a
mass of $173.18 \pm 0.56 {\rm (stat)}  \pm 0.76 {\rm (syst)}$~GeV~\cite{topmassaverage}. Due to its high
mass and its short lifetime the top quark is believed to play a special role in
electroweak symmetry breaking, serves as a window to physics beyond the
standard model (SM), and provides a unique environment to study a bare
quark. 
While the existence of the top quark was predicted by the SM, 
conclusive evidence is still required
that the particle we observe is indeed the one predicted by theory.
In order to use the top quark to search for new physics, it is therefore crucial to precisely determine the production rate and 
properties of top quarks and to confront the results with SM predictions.
In particular, if top quark properties as for example the top width,
the spin correlation or the forward backward charge asymmetry deviate
from the SM prediction, it could indicate physics beyond the SM.

As of today, two colliders with high enough energy exist where top
quarks can be produced. The first one is the Tevatron collider at
Fermilab, which is a proton antiproton collider. During Run~I of the
Tevatron, lasting from 1992 to 1996,
$p\bar{p}$ collisions at 1.8~TeV collision energy were taking
place. In 1995, the CDF and D0 experiments discovered the top quark
with 67~pb$^{-1}$ and 50~pb$^{-1}$ of integrated luminosity,
respectively~\cite{cdfdiscovery, d0discovery}. In 2001, Run~II started
with 1.96~TeV $p\bar{p}$ collisions. Tevatron Run~II lasted until September 30th,
2011, providing approximately 10.5~fb$^{-1}$ of integrated luminosity
per experiment. 
The second collider where top quarks can be
produced~\cite{atlasdiscovery, cmsdiscovery} is the Large
Hadron Collider (LHC) at CERN, where protons are collided with protons
at a center of mass energy of 7~TeV. LHC started operating in 2010,
and has delivered already more than 3~fb$^{-1}$ of collision data to
the two multi-purpose detectors at LHC, ATLAS and CMS. 
Due to the high
center of mass energy, the $t\bar{t}$ production rate at LHC is about
a factor of 20 larger than at the Tevatron, making the LHC a top quark
factory~\cite{dave, francesco}. 
The large 
datasets at all four experiments, ATLAS, CDF, CMS and D0, enable us to
measure top quark production and several top quark properties with
high precision. In the following, recent studies of intrinsic top quark
properties, top production and decay properties and  direct searches
for physics beyond the SM at ATLAS, CDF, CMS and D0 are presented, using up to
5.4~fb$^{-1}$ of data for the Tevatron experiments and up to
1.1~fb$^{-1}$ of data for the LHC experiments.

\section{Top Quark Intrinsic Properties}
Intrinsic properties of the top quark include its mass, its
electric charge, its lifetime and its width. In this section
recent results of the measurement of the top quark mass as well as the
top antitop quark mass difference are discussed.

\subsection{Top Quark Mass}
The mass of the top quark, $m_t$,  is a free parameter of the SM. In
addition to the necessity to determine this free parameter itself, the top quark mass together with the $W$ boson mass 
yields a mass constraint on the so far undiscovered Higgs boson. 

In order to measure the top quark mass as precisely as possible,
several extraction techniques have been developed, which can be
classified into template methods, ideogram methods and the Matrix
Element method. The simplest method is the template method, where
templates are constructed that depend on the top quark mass, which
can then be fitted to data. In the lepton+jets channel, where one of
the $W$ bosons from the top quark decays into a charged lepton and a
neutrino that leaves the detector without interacting,  and the other one into two quarks, the full event kinematics
can be reconstructed using a kinematic fitter by constraining the
invariant mass of the charged lepton and the neutrino  to the known
mass of the $W$ boson. For dileptonic events, the two neutrinos in the
final state cause underconstrained kinematics. In this case,
additional integration over the unknown quantities is required, which
is done using neutrino weighting or matrix weighting techniques. In
the neutrino weighting technique, the pseudorapidities $\eta$~\footnote{The rapidity $y$ and pseudorapidity $\eta$ are defined as functions
    of the polar angle $\theta$ and parameter $\beta$ as
    $y(\theta,\beta) \equiv                                                                                
    {\frac{1}{2}} \ln{[(1+\beta\cos{\theta})/(1-\beta\cos{\theta})]}$ and
    $\eta(\theta) \equiv y(\theta,1)$, where $\beta$ is the ratio of a particle's
    momentum to its energy. } of the two
neutrinos are sampled. For each choice of $\eta$, the kinematics of
the event can be resolved with up to two solutions of
the neutrino transverse momenta. These momenta are then used to
calculate a weight for each solution and each assumed combination   of
$\eta$  values,
based on the agreement between the calculated neutrino transverse
momenta and the  measured value of the missing
transverse energy. In the matrix weighting technique, a top quark mass
dependent weight is
calculated for each event by determining the consistency of the
top-antitop quark momenta, using the assumed top quark mass, 
 with the observed lepton and jet momenta and the
missing transverse energy. 
In case of alljets events, where both $W$ bosons from the top quark
decay into a pair of quarks, the kinematics of the event is fully
determined, but the challenge lies in finding the correct permutation
of jets to match the top and antitop quarks. 

The most precise technique to measure the top quark mass is the
so-called Matrix Element (ME) method. For the ME method, the full
kinemtic information of each event is used, by calculating per-event
signal probabilities $P_{sig}(x;m_t)$ and background probablities $P_{bkg}$, where $x$ denotes the momenta
of the final state partons. The probabilities are calculated by
integrating over the leading
order (LO) matrix element for the $t\bar{t}$ production, folded with
the parton distribution functions and transfer functions. The latter
describe  the transition of
the parton momenta as needed for the matrix element into the measured
momenta of
the final state particles from the top quark decays. 
The measured top
quark mass is then obtained by maximizing the likelihood of the
product of these per-event probabilities. Since only leading order
matrix elements are used and usually the background per-event
probabilities get approximated by only using the matrix element for
the  largest background, the method needs to be calibrated using
ensemble tests.

Finally, a third method that is an approximation of the ME method, is
the so-called ideogram technique. Like in the ME technique, per-event
probabilities are calculated, but with the modification that instead
of matrix elements a kinematic fitter is used. The idea behind this
method is to achieve a statistical uncertainty close to the ME method,
but without the huge computational effort as needed for the ME
technique. 

Independent of the method, the largest contribution to the systematic uncertainty on the top quark
mass comes from the jet energy scale (JES). By constraining the
invariant mass of the two jets coming from the $W$ boson to the known
$W$ boson mass, the JES can be fitted in-situ in the lepton+jets and
alljets channels, resulting
in a reduced dependence of the top quark mass measurement on the JES
uncertainty. In the dilepton final state the in-situ JES fit is not
possible, but CDF recently performed a simultaneous measurement of
$m_t$ in the lepton+jets and dilepton final state, where the
fitted JES from the hadronically decaying $W$ boson in the lepton+jets
channel was applied to the jets in the dilepton final state~\cite{dilepmasscdf}.

During the course of Run~I and Run~II of the Tevatron, all described techniques have been developed,
refined and used to measure the top quark mass as precisely as
possible. Recent measurements of $m_t$ using template techniques are
performed by CDF in the alljets ($m_t=172.5 \pm 2.0 {\rm
  (stat+syst)}$~GeV~\cite{alljetsmasscdf} using 5.8~fb$^{-1}$),
dilepton ($m_t=170.3 \pm 3.7  {\rm (stat+syst)}$~GeV~\cite{dilepmasscdf} using 5.6~fb$^{-1}$) and \met+jets
($m_t=172.3 \pm 2.6  {\rm (stat+syst)}$~GeV~\cite{metjetsmasscdf} using 5.7~fb$^{-1}$) channels, by the ATLAS collaboration in the lepton+jets
final state ($m_t=175.9 \pm 0.9  {\rm (stat)} \pm 2.7  {\rm (syst)}$~GeV~\cite{ljetsmassatlas} using 0.7~fb$^{-1}$) and by
CMS in the dileptonic final state using the matrix weighting technique
($m_t=175.5 \pm 4.6  {\rm (stat)} \pm 4.6  {\rm (syst)}$~GeV~\cite{dilepmasscms} using 36~pb$^{-1}$). Using the ideogram
method, recently CMS measured $m_t=173.1 \pm 2.1  {\rm (stat)}^{+2.8}_{-2.5}  {\rm (syst)}$~GeV in the lepton+jets final
state~\cite{ljetsmasscms} with 36~pb$^{-1}$ of data. New results using
the  ME method are a measurement from D0 in the dileptonic final state
($m_t=174.0 \pm 3.0  {\rm (stat+syst)}$~GeV~\cite{dilepmassd0} using 5.4~fb$^{-1}$) and the lepton+jets
final sate ($m_t=174.9 \pm 1.5  {\rm (stat+syst)}$~GeV~\cite{ljetsmassd0}) as well as a measurement in the
lepton+jets channel by CDF ( ($m_t=173.0 \pm 1.2  {\rm
  (stat+syst)}$~GeV~\cite{ljetsmemasscdf} using
5.6~fb$^{-1}$),  the latter being the single most precise measurement
of the top quark mass to date. 

\begin{figure*}[t]
\centering
\includegraphics[height=70mm]{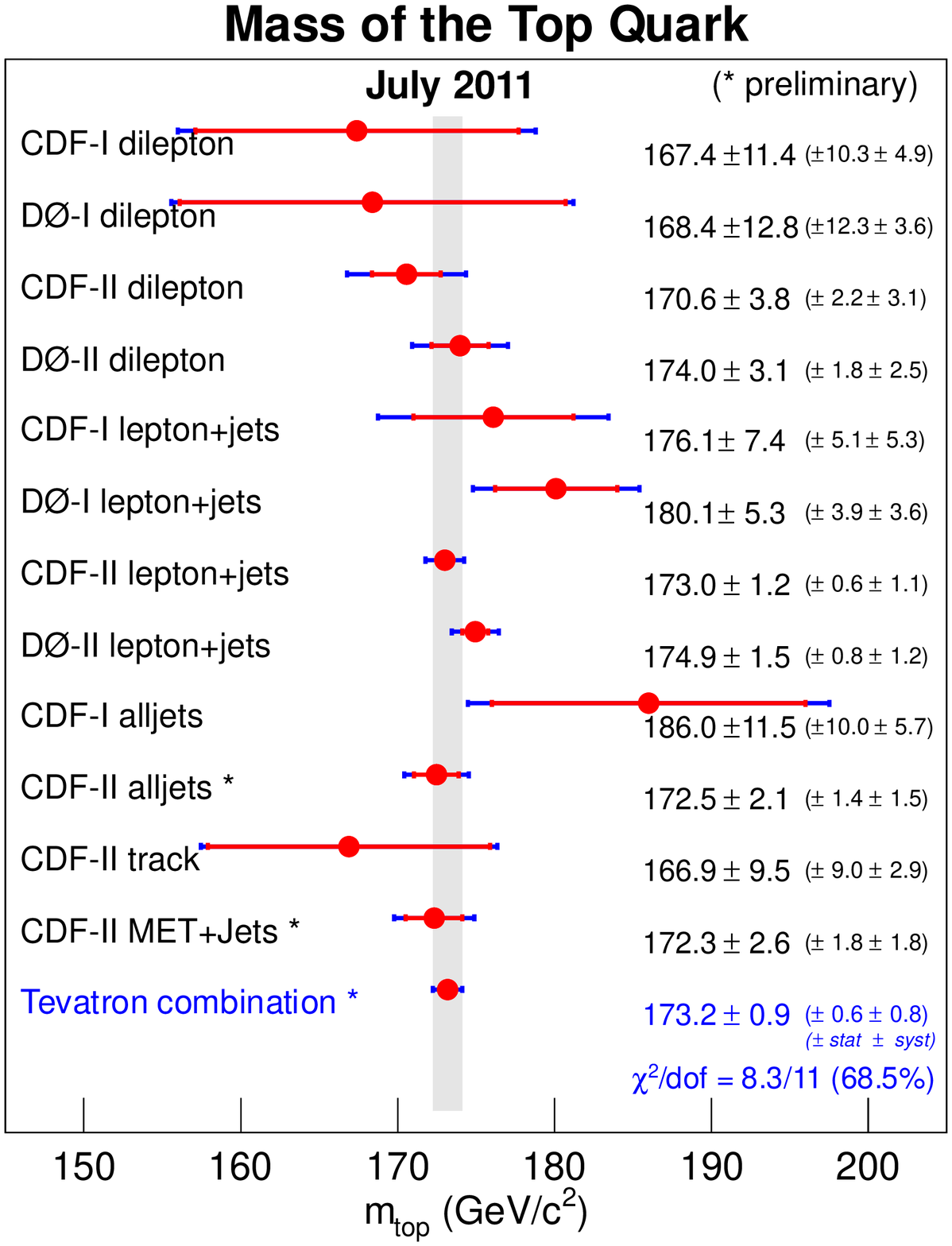}
\includegraphics[height=70mm]{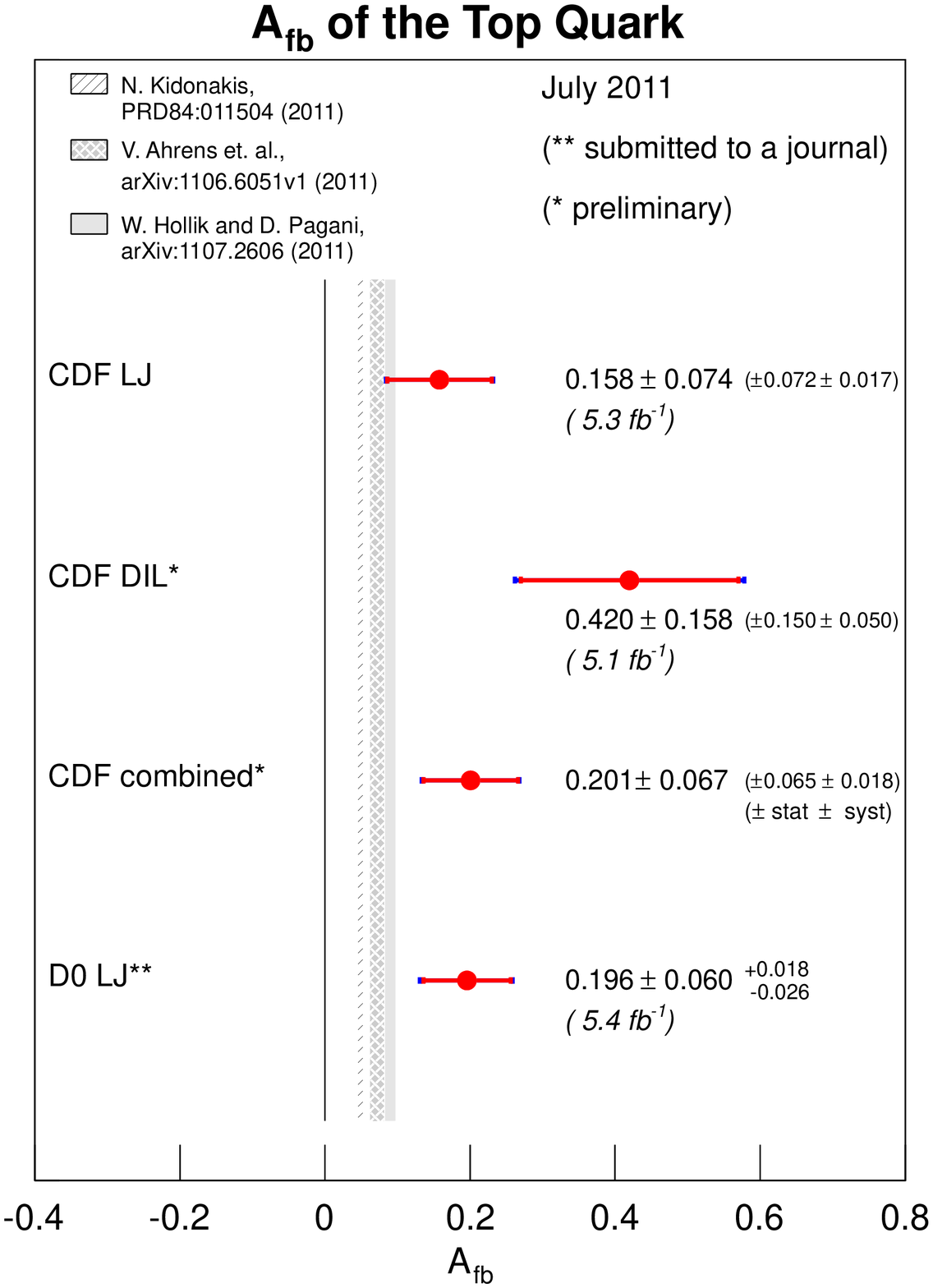}
\caption{Left: Tevatron top quark mass measurements in the different
  final states during Run~I and Run~II and their combination~\cite{topmassaverage}. Right:
  Measurements of the forward backward charge asymmetry $A_{fb}$ at
  the Tevatron~\cite{dilepljetsasymcdf}. } \label{combiplots}
\end{figure*}

A combination of all recent as well as older top
quark mass measurements using the various measurement techniques from Run~I and Run~II of the Tevatron has been
performed, yielding $m_t=173.18 \pm 0.56 {\rm (stat)}  \pm 0.76 {\rm
  (syst)}$~GeV~\cite{topmassaverage}. 
A precision of 0.6\% is achieved on the top quark mass measurement,
which 
is dominated by systematic uncertainties.
The main sources
of systematic uncertainties arise from uncertainites due to the differences of the JES for different jet
flavors  and uncertainties on the signal
modeling. The latter include initial and final state radiation, color reconnections, and
next-to-leading order (NLO) versus LO Monte Carlo (MC) generators,
from uncertainties due to the in-situ fit of the jet energy scale and
related residual dependences of JES on jet $p_T$ and $\eta$. Figure~\ref{combiplots}~(left) shows the different Tevatron top quark mass
measurements and the combination.

All direct top quark mass measurements rely on MC simulations for the template
construction or calibration of the method. These simulations are
perfomed in LO quantum chromodynamics
(QCD), with higher order effects simulated through parton showers at modified leading logarithms (LL) level.
 Since the top quark mass is
a convention-dependent parameter beyond LO QCD, it is important to know how to interpret the result of the
direct measurement in terms of renormalization conventions. Currently,
it is still under theoretical investigations how the measured top
quark mass from MC and the top quark pole or ${\overline {MS}}$ mass are
related. Recently, the D0 Collaboration has determined the top quark
mass from the measurement of the $t\bar{t}$ cross section by comparing
the measured $t\bar{t}$ cross section to inclusive cross section
calculations versus top quark mass, allowing an 
unambiguous interpretation in the pole or ${\overline {MS}}$ mass
scheme~\cite{d0massfromxsec}. Using the pole mass for inclusive cross section calculations
D0 extracted a pole mass of, for example, $m_t=167.5^{+5.2}_{-4.7}$~GeV for the
cross section calculation from Ref.~\cite{mochuwer}. Doing the same
extraction again but with a calculation in the ${\overline {MS}}$ mass
scheme yields about 7~GeV smaller values for $m_t$. Recently, ATLAS~\cite{atlasmassfromxsec}
and CMS~\cite{cmsmassfromxsec} also performed a top quark mass from cross section extraction
in the pole mass scheme, which yields consistent results with the D0 values.

\subsection{Top Antitop Mass Difference}
The direct top quark mass measurements assume the top and the antitop
quark mass to be identical, as required by the SM. If particles and their
corresponding anti-particles would not have equal masses, this would
indicate CPT violation. By dropping the assumption of equal top and
antitop quark masses, the 
CDF, CMS and D0 collaborations have performed measurements of the  top antitop quark
mass difference. The first measurement of  the mass difference between a bare quark
and its antiquark was performed by
the D0 collaboration using the ME method, by extending the event probabilities $P_{sig}(x;m_t)$
to $P_{sig}(x;m_t, m_{\bar{t}})$ using 1~fb$^{-1}$ of
data in the lepton+jets final state~\cite{firstmassdiffd0}. D0
repeated the measurement using 3.6~fb$^{-1}$ of data, resulting in
$m_t-m_{\bar{t}}=0.8 \pm 1.8 {\rm (stat)} \pm 0.5 {\rm
  (syst)}$~GeV~\cite{massdiffd0}, which is consistent with the SM. The
CDF collaboration performed the mass difference measurement using a
template technique in the lepton+jets channel using 5.6~fb$^{-1}$ of data, resulting in $m_t-m_{\bar{t}}=-3.3 \pm 1.4 {\rm (stat)} \pm 1.0 {\rm
  (syst)}$~GeV~\cite{massdiffcdf}. The most precise top antitop quark
mass difference measurement to date was recently performed by the CMS
collaboration by extending the top mass measurement in the muon+jets
channel using the ideogram technique. In particular, the top quark
mass is measured separately in samples with positively and negatively
charged muons, and the resulting top mass values are subtracted. Using 1.09~fb$^{-1}$ of data,
the CMS result is $m_t-m_{\bar{t}}=1.2 \pm 1.2 {\rm (stat)} \pm 0.5 {\rm
  (syst)}$~GeV~\cite{massdiffcms}. All top antitop mass difference
measurements are still dominated by 
statistical uncertainties.

\section{Top Quark Production Properties}
In this section two measurements probing the properties of the top
quark production are discussed, namely the measurement of the $t\bar{t}$ forward backward charge
asymmetry and the $t\bar{t}$ spin correlations. The latter not only
probes the production but the full chain of $t\bar{t}$ production and decay.

\subsection{\boldmath $t\bar{t}$ Forward Backward Charge Asymmetry}
At LO QCD, the production of top antitop quark pairs is forward
backward symmetric in quark antiquark annihilation processes. However,
at higher order interferences between different diagrams cause a
preferred direction of the top and antitop quarks and therefore an
asymmetry. In particular, at NLO, the leading contribution is from the
the interference between tree level
and box diagrams, which yield a positive asymmetry, where the top quark is
preferentially emitted in the direction of the incoming
quark. A deviation from the
SM prediction could indicate physics beyond the SM. For example,
axial currents could contribute to the s-channel production and change
the asymmetry.

Due to the Tevatron being a $p\bar{p}$ collider, where the $t\bar{t}$ production is
dominted by the interaction of a valence quark and a valence antiquark
and therefore the (anti)quark
direction almost always coincides with the direction of the incoing
(anti)proton, the measurement of the forward backward charge asymmetry at
the Tevatron is conceptionally easy. At the LHC, which is a $pp$
collider, two factors result in a more challenging measurement of the
asymmetry as well as in a smaller theory prediction: At 7~TeV,
gluon-gluon fusion dominates the  $t\bar{t}$ production (which
contributes about 85\% to the $t\bar{t}$ production), which does
not cause an asymmetry, and the direction of the incoming quark is
unknown. Therefore, the strategies to measure the asymmetry at the two
colliders are different.

At CDF and D0, the asymmetry is defined in terms of the difference
between the rapidity of the top and antitop quarks, $\Delta y$. The
assignment of the final state particles to top and antitop quarks is  determined by applying  kinematic fitting techniques
to the fully reconstructed $t\bar{t}$ events in the lepton+jets  and
dilepton final states. The charge of the lepton(s) is used to
determine which combination of final state objects belongs to the  top
and which to the  antitop quark. 
The asymmetry is then
defined as $A_{fb}=[N(\Delta y >0) - N(\Delta y<0)]/[N(\Delta y >0) +
N(\Delta y<0)]$, where $N(\Delta y >0)$ and $N(\Delta y <0)$ are the
number of events with rapidity difference larger and smaller zero,
respectively. 
Alternatively, the asymmetry can be extracted from the rapidity
 of the lepton(s) only, which has the advantages
that no complete
    reconstruction of the top and antitop quarks and their decays
is required and that the directions of the charged
leptons can be measured with good resolution. The disadvantage is that
the direction of the lepton is not fully correlated to the top quark
direction, resulting in a loss of sensitivity. 
In order to compare to theory predictions, the measured $t\bar{t}$
forward backward asymmetries are corrected for acceptance and
resolution effects to obtain the inclusive generated asymmetry. The
correction is done using a $4 \times 4$ matrix-inversion at CDF and with
regularized unfolding at D0.

At ATLAS and CMS, the asymmety definition used for the recent
measurements relies on the fact that $t\bar{t}$ production via
$q\bar{q}$ annihilation is dominated by initial valence quarks with
large momentum fractions and initial antiquarks from the sea with
smaller momentum fractions. An asymmetry with the top quark being
preferentially emitted in the direction of the incoming quark 
therefore has the effect that the top quarks has a wider rapidity
distribution than the antitop quarks, which are more
central. Accordingly, the asymmetry is measured using the definition $A_{C}=[N(\Delta |y| >0) - N(\Delta |y|<0)]/[N(\Delta |y| >0) +
N(\Delta |y|<0)]$, with $\Delta |y|$ being the difference of the
absolute rapidity of the top and antitop quark. 

The latest measurements of the asymmetry at the Tevatron were
performed in the lepton+jets channel at CDF and D0, and in the
dilepton final state by CDF. The CDF collaboration measured an
inclusive generated asymmetry of $A_{fb}=0.158\pm 0.074$ using 5.3~fb$^{-1}$ of
data in the lepton+jets channel~\cite{ljetsasymcdf}, and $A_{fb}=0.420 \pm 0.158$ in the
dilepton final state with 5.1~fb$^{-1}$ of
data~\cite{dilepasymcdf}. The combination of these two measurements
results in $A_{fb}=0.201\pm 0.067$~\cite{dilepljetsasymcdf}. The D0 measurement
with 5.4~fb$^{-1}$ of data in the lepton+jets channel yields
$A_{fb}=0.196 \pm 0.060 {\rm (stat)}^{+0.018}_{-0.026} {\rm
  (syst)}$~\cite{ljetsasymd0}. A summary of these results and the
theory predictions is shown in Fig.~\ref{combiplots}~(right). All results are still dominated by
statistical uncertainties. Comparing the measurement to various
theoretical predictions~\cite{asymtheomix} and the prediction of
\mcatnlo~\cite{mcnlo} MC shows about a two sigma
deviation  towards higher
values of the measurements from the prediction. So far it is not clear whether this deviation comes from new
physics contributions or modeling of the SM or anything else, causing
a strong interest in the asymmetry measurements. Various tests to
check the MC modeling have been performed, as for example a test performed  by the
D0 collaboration to check  the sensitivity to the modeling of
the transverse momentum of the $t\bar{t}$ system, $p_T(t\bar{t})$. This test showed
that the  asymmetry predicted by several MC generators is indeed sensitive to
$p_T(t\bar{t})$, which will require further investigations in the
future. 

Recently, the ATLAS and CMS collaborations performed their first
measurement of $A_{C}$ in the lepton+jets final state. Using
0.7~fb$^{-1}$, the ATLAS collaboration measured $A_C=-0.024\pm0.016
{\rm (stat)} \pm 0.023 {\rm (syst)}$, to be compared to a theory
prediction of 0.6\%~\cite{ljetsasymatlas}. The CMS collaboration used
the pseudorapidity instead of the rapidity for the measurement of
$A_C$, resulting in $A_C^{\eta}= -0.016 \pm 0.030 {\rm
  (stat)}^{+0.010}_{-0.019} {\rm (syst)}$ using 1.09~fb$^{-1}$ of
data~\cite{ljetsasymcms}. Neither result shows a significant
deviation from the SM predictions.

Besides the inclusive measurement, it is interesting to investigate the
dependence of the asymmetry on various variables, as for example the
rapidity or the invariant mass of the top antitop quarks,
$m_{t\bar{t}}$. CDF and D0 investigated the $m_{t\bar{t}}$ dependence
by measuring $A_{fb}$ for regions of $m_{t\bar{t}}<450$~GeV and
$m_{t\bar{t}}>450$~GeV. While in D0 data,  no significant dependence
was observed~\cite{ljetsasymd0}, an excess of about three sigma standard deviation
from the \mcatnlo\ prediction was observed by the CDF collaboration for
$m_{t\bar{t}}>450$~GeV~\cite{ljetsasymcdf}. The CMS collaboration also checked the
dependence on $m_{t\bar{t}}$, and no significant dependence showed up
in the CMS data~\cite{ljetsasymcms}.

\subsection{\boldmath $t\bar{t}$ Spin Correlations}
While the top quarks are produced unpolarized at
hadron colliders, the spins of the top and antitop quarks are expected
to be correlated. Due to the short lifetime of the top quark, which is
shorter than the time scale for hadronization, the information of the top quark's spin is preserved in
its decay products, enabling the measurement of the spin correlation
of the 
top and antitop quark in $t\bar{t}$ events. Recently, two different
methods have been explored to measure $t\bar{t}$ spin correlations,
namely a template based method relying on angular distributions, and a
matrix element based  method.

Template based methods were used by the ATLAS, CDF and D0
collaborations. At D0 and CDF, the measurements are based on the fact
that the doubly differential cross section, $1/\sigma \times d^2
\sigma /(d \cos \theta_1 d \cos \theta_2)$ can be written as $1/4
\times 
(1-C \cos \theta_1 \cos \theta_2)$, where $C$ is the spin correlation
strength, and $\theta_1$ ($\theta_2$) is the angle of the down-type
fermion from the $W^{+}$  ($W^{-}$) boson or top (antitop) quark decay in the top
(antitop) quark rest frame with respect to a quantization axis. Common
choices are the helicity basis, where the quantization axis is the flight direction of the top
(antitop) quark in the $t\bar{t}$ rest frame, and the beam basis, where the
quantization axis is the beam axis. A third common choice of
quantization axis is the off-diagonal basis, which yields the helicity
axis for ultra-high energy and the beam axis at threshold. 
The SM prediction for the spin correlation strength $C$
depends on the collision energy and the choice of quantization axis,
and is $C=0.78$ for the Tevatron in the beam basis at NLO~\cite{bernreutherspin}. 
 The spin correlation strength $C$ can be presented as the number
of events where top and antitop have the same spin direction minus the number of events with
opposite spin direction, normalized to the total number of $t\bar{t}$ events, multiplied with a factor
representing the analyzing power of the down-type fermion used to
calculate the angles. The latter factor is one for leptons and
down-type quarks from the $W$ boson decay at LO QCD, but smaller for
up-type quarks and the $b$-quark from  top quark decay. Since it is
experimentally challenging to distinguish up-type from down-type
quarks, the dilepton channel is best to perform spin correlation
measurements. The CDF and D0 collaborations  performed an analysis of
the spin correlation strength $C$ by fitting templates for $C=0$ and
the SM value of $C$ of the distribution  $\cos
\theta_1 \cos \theta_2$ to data. Using 2.8~fb$^{-1}$ at CDF and
5.4~fb$^{-1}$ at D0, the measurement of $C$ in the beam basis yields
$C=0.32^{+0.55}_{-0.78} {\rm (stat+syst)}$~\cite{cdfdilepspin} and
$C=0.10 \pm 0.45 {\rm (stat+syst)}$~\cite{d0dilepspin}, in agreement
with SM prediction. Similar to these two analyses in the dilepton final
state, CDF performed the first extraction of $t\bar{t}$ spin
correlations by fitting templates of equal and opposite $t\bar{t}$
helicity  to data. The measured quantity is then translated into
$C$. Using a dataset of 4.3~fb$^{-1}$, CDF measured $C=0.72 \pm 0.64
{\rm (stat)} \pm 0.26 {\rm (syst)}$ in the beam basis~\cite{cdfljetsspin}.

Recently, the D0 collaboration explored a different method to measure
$t\bar{t}$ spin correlation, where per-event signal probabilities
$P_{sig}(H)$ are calculated using matrix elements that include
spin correlations ($H=c$) and do not include spin correlations
($H=u$), 
and are translated into a discriminant
$R=P_{sig}(H=c)/[P_{sig}(H=c)+P_{sig}(H=u)]$~\cite{melnikovschulze}. Applying
this technique to the same D0 dataset of 5.4~fb$^{-1}$ of dilepton events
as for the template based method, results in a 30\% improved
sensitivity, yielding $C=0.57 \pm 0.31 {\rm
  (stat+syst)}$~\cite{d0dilepmespin}. The matrix element-based method
has been extended to the lepton+jets final state using 5.3~fb$^{-1}$
of D0 data, and by combining the
measurement in 
dilepton and  lepton+jets events first evidence for spin correlation
was reported recently~\cite{d0ljetsdilepmespin}.
All Tevatron measurements are in
  agreement with the NLO SM prediction, and all are still limited by
  statistics.  

While at the Tevatron $q\bar{q}$ production at threshold dominates, the dominant $t\bar{t}$ production at LHC
is via gluon-gluon fusion. In a recent theory article~\cite{parkemahlon}
it was suggested that the $t\bar{t}$  production at the LHC at low parton
collision energy is dominated by fusion of like helicity gluons. A
simple variable was proposed, which is the difference in azimuthal
angle of the two leptons in the lab frame, $\Delta \phi$, in dileptonic final
states. In contrast to the discussed template based measurements at CDF
and D0, no full reconstruction of the $t\bar{t}$ system is required
and $\Delta \phi$ can therefore be precisely measured. The ATLAS collaboration
performed the first measurement of $t\bar{t}$ spin correlation at the
LHC using the variable $\Delta \phi$, which yields
$C_{heli}=0.34^{+0.15}_{-0.11} {\rm (stat+syst)}$ in the helicity basis
using 0.7~fb$^{-1}$ of data~\cite{atlasspin}. This is in agreement
with the NLO SM prediction of $C_{heli}^{theo}=0.32$. The measurement
by ATLAS is dominated by systematic uncertainties due to the signal
modeling, mainly the modeling of initial and final state radiation.

\section{Top Quark Decay Properties}
Since physics beyond the SM could also show up in top quark decays, it is
important to study top quark decay properties, as for example the
ratio of branching fractions, the $W$ helicity or anomalous
couplings. Recent studies of these properties are presented in this section.

\subsection{Ratio of Branching Fractions}
In the SM, the top quark decays to a $W$ boson and a $b$-quark with
almost 100\% probability. In the measurement of the ratio of branching
fractions $R= B(t \rightarrow Wb)/B(t \rightarrow Wq)$, with $q= b, s,
d$, the possibility is studied that the quark from the top quark decay
can be a light down-type quark. Physics beyond the SM or a fourth
generation of quarks could cause $R$ to be below its SM value of one.
In a new measurement of $R$ by the
D0 collaboration, using 5.4~fb$^{-1}$ of data, the ratio of branching
fractions has been measured in the dilepton and lepton+jets final
states~\cite{d0rb}. In the lepton+jets channel, the distribution of
events with $0$, $1$ or $\ge2$ identified $b$-jets is used to
discriminate $R$, while in
the dilepton channel the distribution of the output of the $NN$
$b$-tagging algorithm~\cite{nnbtag} is analysed. To reduce the
influence of systematic uncertainties that change the normalization of
signal and background contributions, the $t\bar{t}$ cross section is
fitted simultaneously with $R$, resulting in $R=0.90 \pm 0.04 {\rm
  (stat+syst)}$, which is the most precise determination of $R$ to
date. Lower limits on $|V_{tb}|$ assuming unitarity of the $3 \times 3$
CKM matrix can then be extracted from $R$, resulting in $|V_{tb}|=0.95
\pm 0.02 {\rm (stat+syst)}$. The measurement is limited by systematic
uncertainties, with the main source coming from uncertainties on
$b$-jet identification. The measured value of $R$ shows about a two
sigma standard deviation from the SM.

\subsection{\boldmath $W$ Helicity and Anomalous Couplings}
The relative orientation of the spin of the $b$-quark and the $W$
boson from the top quark decay are constrained in the SM by the fact
that $W$ bosons couple purely left-handed to fermions. The fractions
of negative ($f_{-}$), zero ($f_0$) and positive ($f_{+}$) helicity of
the $W$ boson are predicted to be $f_{-}=0.685\pm0.005$,
$f_0=0.311\pm0.005$ and $f_{+}=0.0017\pm0.0001$ at
next-to-next-to-leading order (NNLO) QCD~\cite{nnloheli}.
Similar to the measurement of the top quark mass and $t\bar{t}$ spin
correlations, various methods can be used to measure the $W$ boson
helicity fractions, namely a template based method and the ME
technique. In the template method, the angle $\theta^{*}$ between the
down-type decay product of the $W$ boson and the top quark in the $W$ boson 
rest frame is measured, which differs for the three possible helicity
fractions, and distributions of the cosine of this angle
are fitted to data. To keep the analysis as model-independent as
possible, the fractions $f_0$ and $f_{+}$ are fitted simultaneously,
only constraining the sum of all three fractions to be one. The CDF
collaboration also uses the ME method to measure the $W$ boson
helicity, where the per-event signal probabilities $P_{sig}$ are
calculated as function of $f_0$ and $f_{+}$ instead of $m_t$, with the
latter fixed to $175$~GeV. Recently, a combination of a D0 measurement
in the dilepton and lepton+jets channel using 5.4~fb$^{-1}$, a CDF
measurement in the lepton+jets final state using 2.7~fb$^{-1}$, and a
CDF analysis in the dilepton final state using 5.1~fb$^{-1}$ has been performed~\cite{whelitevcombi}.
Using the model independent approach, the combination yields
$f_0=0.732 \pm 0.063 {\rm (stat)} \pm 0.052 {\rm (syst)}$ and
$f_{+}=-0.039 \pm 0.034 {\rm (stat)} \pm 0.030 {\rm (syst)}$, in good
  agreement with the SM prediction.  Furthermore, the CDF collaboration
  updated their measurement in the dilepton final state using
  5.1~fb$^{-1}$, improving the sensitivity by applying $b$-jet
  identification. This result is not yet included in the Tevatron
  combination and yields  $f_0=0.71^{+0.18}_{-0.17} {\rm (stat)} \pm
  0.06 {\rm (syst)}$ and
$f_{+}=-0.07 \pm 0.09 {\rm (stat)} \pm 0.04 {\rm (syst)}$~\cite{cdfdilepnew}.

Using the template fit of the $\cos \theta^{*}$ distributions, the ATLAS
collaboration performed the first measurement of the $W$ helicity fractions
 at the LHC. In the lepton+jets channel, the model independent fit
is applied, resulting in $f_0=0.57 \pm 0.07 {\rm (stat)} \pm 0.09 {\rm (syst)}$ and
$f_{+}=0.09 \pm 0.04 {\rm (stat)} \pm 0.08 {\rm
  (syst)}$ with 0.70~fb$^{-1}$ of data~\cite{atlaswheli}, where the
dominant systematic uncertainties are from signal modeling and
modeling of initial and final state radiation. For the
dilepton final state the data sample was too small to perform the model
independent approach, and instead $f_{+}$ is fixed to zero and $f_0$ is
fitted. The dilepton and lepton+jets combined measurement with
$f_{+}=0$ yields $f_{0}=0.75 \pm 0.08 {\rm (stat+syst)}$, which is
consistent with the SM prediction.

Using the measurements of the $W$ helicity fractions, or  single top
events, constraints on anomalous couplings of the $Wtb$ vertex can be
extracted. In the SM only left-handed vector couplings ($V_L$) are allowed,
while a more general $Wtb$-Lagrangian could allow right-handed vector ($V_R$),
left-handed tensor ($g_L$) or right-handed tensor ($g_R$) couplings. 
The ATLAS collaboration extracted limits on $V_R$, $g_L$ and $g_R$
using the $W$ helicity measurement~\cite{atlaswheli}, while the D0
collaboration extracted limits on these couplings using the selection
for single top events~\cite{anomalouscouplingsnewd0}. A
previous analysis performed by the D0 collaboration on up to 2.7~fb$^{-1}$ of data  showed that the combination
of $W$ helicity and single top measurements improves the sensitivity
to all three anomalous couplings wrt. the single measurements~\cite{anomalouscouplingsoldd0}.

\section{Other Top Quark Studies and Searches}
Precision measurements of top quark properties and their comparison to
SM predictions are crucial to study the nature of the particle
discovered in 1995 and believed to be the top quark. Additionally,
direct searches in the top quark sector could reveal physics beyond
the SM or an unexpected behaviour of the top quark. In the following,
an example of a study of color flow in $t\bar{t}$ events and a search
for flavor changing neutral currents (FCNC) are presented.

\subsection{\boldmath Color Flow in $t\bar{t}$ Events}
In QCD, color charge is a conserved quantity, causing two final-state
particles on the same color-flow line to be color connected to each
other. In a recent paper~\cite{theorypull}, a tool called jet pull,
which is related to the jet energy pattern in the $\eta-\phi$ plane, has been suggested to measure color
flow between a jet pair  and distinguish color-octet from
color-singlet states. For jets from color singlet states, such as  a $b$-jet pair from Higgs boson decay, the pulls of the jets tend to point
towards each other, while for a $b$-jet pair from a color-octet gluon the
pulls would point in opposite directions along the collision
axis. Before such a tool can be applied to new physics searches, it
has to be studied in  a known environment, as for example in
semileptonic $t\bar{t}$ events, where the two light jets from the $W$
boson decay are expected to come from a color-singlet. 
Recently, the D0 collaboration performed the first study of color flow
using the jet pull variable in $t\bar{t}$ events. 
Using 5.3~fb$^{-1}$ of data, a SM $t\bar{t}$ MC with a color-singlet
$W$ boson has been compared to a $t\bar{t}$ sample with a
hypothetical, hadronically decaying, 
color-octet ``W'' boson. The jet pull variable has been applied to  
extract the fraction $f$ of color-singlet hadronic $W$ boson decays,
resulting in $f=0.56 \pm 0.38 {\rm (stat+syst)} \pm 0.19 {\rm (MC
  stat)}$~\cite{colorflowd0}, with an expected 99\% confidence level
exclusion of the
color-octet ``W'' boson. This study is still dominated by the
statistics of the analyzed data sample.

\subsection{Search for Flavor Changing Neutral Currents}
Transitions between quarks of the same electric charge but different
flavors could occur if FCNCs exist. In the SM, FCNC are
suppressed. Therefore, observation of FCNC would indicate physics
beyond the SM, as for example supersymmetric models or quark
compositeness. At the Tevatron, various searches for FCNC in the top
sector have been performed, for example searches for the decay
$t\rightarrow gq$, with $q=c, u$ in single top
events~\cite{singletopfcnc} and searches for  $t\rightarrow Zq$ in
$t\bar{t}$ events. For the latter, a search in dileptonic $t\bar{t}$
events has been performed by the CDF collaboration, using
1.9~fb$^{-1}$ of data~\cite{cdffcnc}, as well as a recent search by
the D0 collaboration using trilepton final states~\cite{d0fcnc}. For
the search by D0 a $t\bar{t}$ sample of 4.1~fb$^{-1}$ of integrated
luminosity  has been used, where at least one of the top quarks decays
to a $Z$ boson and a light quark, with the $W$ boson and  $Z$ boson(s) decaying
leptonically, resulting in at least three leptons in the final
state. To increase the sensitivity, distributions sensitive to FCNC,
like the scalar sum of the transverse momenta of all leptons, jets,
and \met, are included in the limit setting procedure. 
This search yields the world's best limits on FCNC to date, of
$B(t \rightarrow Zq)<3.2\%$ at 95\% confidence level.

Using 35~pb$^{-1}$ of data, ATLAS performed a search simiar to the
trilepton search by D0, but with counting events
only~\cite{atlasfcnc}, resulting in $B(t \rightarrow Zq)<17\%$ at 95\%
confidence level. Furthermore, ATLAS searched for anomalous single top production
through FCNC using the same dataset. Neural networks are used to
separate $gq \rightarrow t$ signal from background, resulting in upper
limits on the cross section $\sigma_{gq \rightarrow t} \times B(t
\rightarrow bW)$ of 17.3~pb at 95\% confidence level.

\section{Conclusion and Outlook}
A collection of the most recent measurements of top quark properties
at the ATLAS, CDF, CMS, and D0 collaborations has been discussed in
this presentation. About 10.5~fb$^{-1}$ of data have been collected by
    the CDF and D0 collaborations in Run~II of the Tevatron, which
    ended on September 30th, 2011. Only about half of this dataset has
    been used so far for top quark studies. 
The Tevatron experiments plan to analyse
the final dataset for those measurement which are
complementary or competitive to  the LHC results, including the top quark mass measurement, the measurement of the
    forward-backward charge asymmetry and $t\bar{t}$ spin
    correlations. With the start of the LHC in 2010, a top quark
    factory was opened. Having collected more than
    3~fb$^{-1}$ already, the comparison of top quark measurements to theory
    predictions reaches new levels of precision. Given the top quark's
    special role in particle physics, it is and will stay an
    interesting particle to study at all four experiments.

\bigskip 
\begin{acknowledgments}
I would like to thank my collaborators from the ATLAS, CDF, CMS and D0
collaborations for their help in preparing the presentation and this
article. I also thank the staffs at Fermilab and CERN and
collaborating institutions, and acknowledge the support from STFC.
\end{acknowledgments}

\bigskip 
\bibliography{basename of .bib file}

\end{document}